\documentclass[12pt,twoside]{article}

\def\TheAuthor{Fabian Hassler}                                                             
\def\TheTitle{Majorana Qubits\footnotemark}                        
\def\TheHeading{Majorana Qubits}                                                                 
\def\TheInstitute{Institute for Quantum Information}                                         
\def\TheAddress{RWTH Aachen University}                           
\def\ThePart{B}                                                                                  
\def\TheChapter{2}                                                                            

\usepackage{ferienschule_12}                                  

\newcommand\pdag{{\phantom\dag}}

\begin{document}

\MakeTitel           

\tableofcontents     

\footnotetext{Lecture Notes of the $44^{{\rm th}}$ IFF Spring
School ``Quantum Information Processing''
(Forschungszentrum J{\"{u}}lich, 2013).  All rights reserved. }

\newpage


\section{Introduction}

Ideas of topology a branch of mathematics which studies whether two
objects can be transformed into each other under continuous deformations
have proven to provide very fruitful concepts also in physics. Examples
in classical physics involve vortices in fluid dynamics, electrical
charges in electrodynamics, and the classification of defects in
crystals \cite{mermin:79}. In quantum mechanics the prime example of a
topological effect is the Aharonov-Bohm effect \cite{aharonov:59}. Soon
after the discovery of the quantum Hall effect \cite{klitzing:80}, it was
realized that the new phase is not characterized by a broken symmetry
but that the topology of the Bloch wavefunction in the Brillouin zone
described by the Chern number changes when entering the quantum Hall
regime \cite{thouless:82}. For some time the quantum Hall effect was
considered the only example of a topological phase in a noninteracting
system, nowadays called topological insulator.  More recently, a second
example was devised theoretically \cite{haldane:87,kane:05,bernevig:06}
in the form of the quantum spin Hall effect and confirmed experimentally in
HgTe quantum wells \cite{konig:07}.  This sparked further research in this
area and culminated in the classification of all topological insulators
and superconductors \cite{schnyder:08,kitaev:09}. The subject of this
lecture are states, called Majorana fermions, which appear at defects in
topological superconductors with broken time-reversal and spin-rotation
symmetry in one or two dimensions.

The outline of the lecture is as follows.  We will first introduce the
notion of Majorana fermions. We will then show how these states appear as
zero energy solutions of the Bogoliubov-de Gennes equation describing a
spinless \emph{p}-wave superconductor in one dimension. We will describe
their usefulness in term of quantum information applications as they
encode quantum information in a protected fashion before we will finish off
discussing a possible experimental implementation. There are by now several
reviews on this subject, see Refs.~\cite{beenakker:11,alicea:12,leijnse:12},
where further information on this subject can be found.

\section{Majorana fermions}

Majorana fermions are fermionic particles which are their own
antiparticles.  Why the corresponding context is quite natural for
bosons---most bosons (phonon, photons, magnons, plasmons, \dots) are their
own antiparticles---this is a rather uncommon property for fermions. In fact
so far no elementary fermionic particle has been experimentally confirmed to
be its own antiparticle. In more mathematical terms, a Majorana operator
$\gamma_1$ (an operator which creates a Majorana particle) is a Hermitian
operator $\gamma_1 = \gamma_1^\dag$ which anticommutes with other Majorana
operators $\gamma_2$ and squares to one $\gamma_1^2=1$; summarizing, the
Majorana operators form a Clifford algebra defined by the anticommutation
relation
\begin{equation}\label{eq:clifford}
  \{ \gamma_k, \gamma_l \} = 2 \delta_{kl}.
\end{equation}
Given the fact that these particles do not exist as elementary particles, we
would like to know how to construct them from conventional Dirac fermions
created by the operator $c^\dag_k$. In fact, it is an easy exercise
in algebra to show that given a set of $N$ Dirac fermions defined by
$c^\dag_k$, $k = 1,\dots,N$, with the canonical anticommutation relations
$\{c_k,c^\dag_l\} = \delta_{kl}$ and $\{c_k, c_l\}=0$, we can construct $2N$
Majorana operators $\gamma_{k}$ via
\begin{align}\label{eq:majorana}
  \gamma_{2k-1} &= c^\pdag_k + c^\dag_k, &
  \gamma_{2k} &= i(c^\dag_k - c^\pdag_k).
\end{align}
Inverting the defining Eq.~\eqref{eq:majorana}, we find an expression of
the Dirac fermions in terms of the Majorana operators
\begin{align}\label{eq:dirac}
  c_k &= \frac12 (\gamma_{2k-1} + i \gamma_{2k}), & 
  c^\dag_k & =\frac12 (\gamma_{2k-1} - i \gamma_{2k}).
\end{align}
The Hilbert space of a single fermionic mode is two-dimensional: the
mode is either filled or empty distinguished by the eigenvalue of the
number operators $n_k = c^\dag_k c^\pdag_k$ which has eigenvalues 0 or
1.\footnote{Note that $n_k$ is idempotent as $n_k^2= c^\dag_k c^\pdag_k
c^\dag_k c^\pdag_k = c^\dag_k (1-c^\dag_k c^\pdag_k ) c^\pdag_k = n_k $
which proves the fact that the eigenvalues of $n_k$ are 0 or 1.
} An operator which will turn out to be important in the following discussion
is the fermion parity operator $\mathcal{P}_k = 1- 2 n_k = (-1)^{n_k}$
which has the eigenvalue $+1$ if the number of fermions is even and $-1$
if the number of fermions is odd. In terms of the Majorana operators,
the parity operator assumes the simple form
\begin{equation}\label{eq:parity}
  \mathcal{P}_k = -i \gamma_{2k-1} \gamma_{2k}.
\end{equation}

So far, the introduction of Majorana fermions was an algebraic trick
to go from one set of complex operators $c_k$ to an equivalent description
in terms of the Hermitian operators $\gamma_{k}$.
Naturally, the question arises if these operators are `physical'
in the sense that they describe the excitations of a physical
system/Hamiltonian. The answer to this question is (maybe surprisingly)
yes: over a decade ago, Kitaev constructed a model which leads to
Majorana fermions \cite{kitaev:01}.

To appreciate the difficulty in constructing such a model as well as to
understand the resolution, we dwell a bit on the requirement/hurdles
to construct such a model. Starting from Dirac fermions, we see from
\eqref{eq:majorana} that the Majorana operators are superposition of
electron and hole operators. We know that ordinary (many-body) quantum
mechanics is invariant under global $U(1)$ transformations $c_k \mapsto
U^\dag_\varphi c_k U_\varphi^\pdag = e^{i \varphi} c_k$ with $\varphi$
an arbitrary phase. The reason for this is the conservation of the total
particle number (or charge for that matter). However, it is easy to check
that under the same transformation Majorana operators corresponding to
the same fermionic mode mix with each other
\begin{equation}\label{eq:gauge_gamma}
  \gamma_{2k-1} \mapsto U_\varphi^\dag \gamma_{2k-1} U_\varphi^\pdag
  = \cos(\varphi) \, \gamma_{2k-1} - \sin(\varphi) \, \gamma_{2k}.
\end{equation}
Thus, if we were to construct a Hamiltonian which has $\gamma_{2k-1}$
(or $\gamma_{2k}$ for that matter) as an elementary, localized excitation,
we will have to break the global $U(1)$ invariance as it mixes the two
different physical modes $\gamma_{2k-1}$ and $\gamma_{2k}$. In fact,
the $U(1)$ symmetry (given by the phase $\varphi$) is broken down to
a $\mathbb{Z}_2$ symmetry (corresponding to $\varphi=0,\pi$).  This is
exactly what happens in superconducting systems, so we should look a bit
more closely into the theory of superconductivity.

\subsection{Bogoliubov-de Gennes equation}

Superconductivity is an ordering phenomena which happens in interacting
system at low temperatures. Experimentally, the state is characterized by a
vanishing of the resistance and, more importantly, by a perfect diamagnetic
response called Mei{\ss}ner-Ochsenfeld effect \cite{gennes,tinkham}.

\pagebreak

Starting from a model of interacting spinless electrons\footnote{As we
want to end up with single unpaired Majorana fermions, we have to get rid
of all possible degeneracies in particular the spin degeneracy.}
\begin{equation}\label{eq:ham_int_e}
  H =  \int\!d^3r\,\left[\frac{\hbar^2}{2m}|\nabla\psi(\bm r)|^2 -\mu |\psi(\bm
  r)|^2\right] 
  -  \frac12 \int\!d^3r\,d^3r'\,
  \psi^\dag(\bm r) \psi^\dag(\bm r')
  V(\bm r - \bm r')\psi(\bm r') \psi(\bm r),
\end{equation}
we employ the mean-field decoupling with the superconducting pair-potential
$\Delta(\bm r-\bm r') = V(\bm r - \bm r') \langle \psi(\bm r') \psi(\bm r)
\rangle$ \cite{negele} and arrive at the effective BCS mean-field Hamiltonian
\begin{multline}\label{eq:ham_eff}
  H_\text{MF} = \int\!d^3r\,\left[\frac{\hbar^2}{2m}
  |\nabla\psi(\bm r)|^2 -\mu |\psi(\bm r)|^2\right] 
  - \frac12 \int\! d^3r\,d^3r'\,  \psi^\dag(\bm r) \psi^\dag(\bm r')
  \Delta(\bm r- \bm r') \\
  - \frac12 \int\! d^3r\,d^3r'\, \Delta^*(\bm r- \bm r') 
  \psi(\bm r') \psi(\bm r) 
  + \frac12 \int\! d^3r\,d^3r'\, \frac{\Delta(\bm r -\bm r')}{V(\bm r-\bm
  r')}.
\end{multline}
In the resulting Hamiltonian, the $U(1)$ degree of freedom $\psi
\mapsto e^{i\varphi} \psi$ for fixed `external field' $\Delta$ is broken
down to a $\mathbb{Z}_2$ degree of freedom just as we wished.

Apart from an unimportant constant, the resulting Hamiltonian can be written
as a quadratic form
\begin{equation}\label{eq:nambu}
  H_\text{MF} =
  \frac12 \int\!d^3r\,d^3r'\,\Psi^\dag(\bm r) h_\text{BdG}(\bm r - \bm r')
  \Psi(\bm r') + \text{const.}
\end{equation}
in Nambu space $\Psi^\dag = \begin{pmatrix}\psi^\dag&\psi\end{pmatrix}$ 
with the Bogoliubov-de Gennes Hamiltonian
\begin{equation}\label{eq:h_BdG}
  h_\text{BdG}(\bm r- \bm r') =  \begin{pmatrix}
    \xi(\bm p) \delta(\bm r -\bm r') &  -\Delta(\bm r - \bm r') \\
    \Delta^*(\bm r - \bm r') &-\xi(\bm p) \delta(\bm r -\bm r')
  \end{pmatrix},
\end{equation}
where $\xi(\bm p ) = \frac{p^2}{2m} - \mu = -\frac{\hbar^2}{2m}\nabla^2 - \mu$;
note that $h_\text{BdG}$ is Hermitian due to the fact that $\Delta(-\bm r)
= - \Delta(\bm r)$.  The Hamiltonian is `diagonalized' by a Bogoliubov
transformation, i.e., by introducing new fermionic operators $\beta_{E_n}$
which fulfill the canonical anticommutation relation and in terms of which
the mean-field Hamiltonian assumes the form \cite{gennes,tinkham}
\begin{equation}\label{eq:ham_mf_bo}
  H_\text{MF} = \frac12 \sum_{E_n} E_n \beta^\dag_{E_n} \beta_{E_n}^\pdag
  +\text{const.};
\end{equation}
here, $E_n$ are the eigenvalues of $h_\text{BdG}$ and $\beta_{E_n} =
\int\!d^3r\,v^\dag_{E_n}(\bm r) \Psi(\bm r)$ where $v_{E_n}(\bm r)$ are
the associated eigenvectors.

Note that in getting to Eq.~\eqref{eq:h_BdG}, we have apparently doubled the
degrees of freedom.  However, the resulting Bogoliubov-de Gennes Hamiltonian
$h_\text{BdG}$ enjoys an additional symmetry: in fact, the particle-hole
symmetry operator $\mathcal{C} = \tau^x K$ with $K$ the complex conjugation
and $\tau^x$ acting on the Nambu index anticommutes with the Hamiltonian
$\{h_\text{BdG}, \mathcal{C}\}=0$. This symmetry guarantees that for every
eigenvector $v_{E_n}$ of $h_\text{BdG}$ to eigenvalue $E_n\geq 0$ there is
an additional eigenvector $v_{-E_n}=\mathcal{C} v_{E_n}$ to eigenvalues
$-E_n$. Expressing this fact in terms of the second quantized Bogoliubov
operators, we have
\begin{equation}\label{eq:bogoliubov}
  \beta^\dag_{-E_n} =  \int\!d^3r\, \Psi^\dag(\bm r) 
  \underbrace{v_{-E_n}(\bm r)}_{\mathcal{C} v_{E_n}} 
  =  \int\!d^3r\, \underbrace{\Psi^\dag(\bm r) 
  \tau^x}_{ \Psi^T(\bm r)} v^*_{E_n}(\bm r)  
  = \beta^\pdag_{E_n}.
\end{equation}
In the end, combining the terms with $E_n$ and $-E_n$, we have $E_n
(\beta^\dag_{E_n} \beta^\pdag_{E_n} - \beta^\dag_{-E_n} \beta^\pdag_{-E_n})
= 2E_n(2\beta^\dag_{E_n}\beta^\pdag_{E_n} -\tfrac12)$ such that we only need to
include the eigenvectors to positive eigenvalues in \eqref{eq:ham_mf_bo}
on the expense of the factor $\tfrac12$ in front of the sum.

Now, we are very close to our goal of realizing Majorana fermions starting
from conventional Dirac fermions. Looking at Eq.~\eqref{eq:bogoliubov},
we see that if an eigenstate $n=0$ of the Bogoliubov-de Gennes Hamiltonian
has a vanishing eigenvalue, $E_0=0$, it is in fact a Majorana fermion with
$\gamma_1= \beta_{0}$.\footnote{Here, it is important that there is only a
single zero mode present. Having two modes $\beta_1$ and $\beta_2$ at zero
energy, we can only conclude that $\beta_1 = \beta_2^\dag$.} Given this
insight, we try to construct a physical situation where \eqref{eq:h_BdG}
incorporates such a zero mode. Here, I want to point out that there is a
principle difference between the more general term Majorana fermion and
the Majorana zero mode. Whereas Majorana fermion simply denotes a `real'
fermion, Majorana zero mode denotes Majorana fermions bound to zero energy
at a topological defect in a superconductor/superfluid.\footnote{However in
the recent literature, the general term `Majorana fermion' is often used to
denote the more special term `Majorana zero mode'.} While the statistics
of the former is simply fermionic, the latter shows non-Abelian exchange
statistics, see below.

\subsection{Spinless \emph{p}-wave superconducting
nanowire}\label{sec:spinless}

Following Kitaev \cite{kitaev:01}, we would like to construct a simple
model which shows Majorana zero modes. Thus, we consider a one-dimensional
situation with $\bm r =z$. The simplest choice of paring, \emph{s}-wave
pairing, with $\Delta(z) = \Delta \, \delta(z)$ is not allowed for
spinless electrons as $\Delta(-z) \neq -\Delta(z)$.  Thus, we take the next
term in the gradient expansion into account, \emph{p}-wave pairing, with
$\Delta(z) = -i \Delta \,\lambda_F\,\delta'(z)$; here, we
have introduced the (reduced) Fermi wavelength $\lambda_F = \hbar/\sqrt{2m
\mu}$ such that $\Delta$ has the dimension of energy. Going over to momentum
space, the Bogoliubov-de Gennes Hamiltonian can be written as
\begin{equation}\label{eq:h_BdG_p_wave} 
  h_\text{BdG} = \xi(p) \tau^z - \Delta \frac{p}{p_F} \,\tau^x.
\end{equation}
In the following, we will assume $\Delta >0$ for convenience.\footnote{The
phase of $\Delta$ is in fact an unobservable quantity as only
phase-differences are observable, e.g., via the Josephson effect
\cite{gennes,tinkham,likharev:79}.} It turns out that the model of the
spinless \emph{p}-wave superconducting wire \eqref{eq:h_BdG_p_wave} is closely
related to the so-called Su, Schrieffer, Heeger model studied some time ago as
a model for polyacetylene \cite{su:79,heeger:88}.  We will not go into a
detailed discussion of the similarities and difference of the two model, we
just want to point out that Su \emph{et al.} have found that their model in
polyacetylene generates zero energy state of certain topological criteria are
satisfied.

\begin{figure}[tb]
  \centering
  \includegraphics[width=0.9\linewidth]{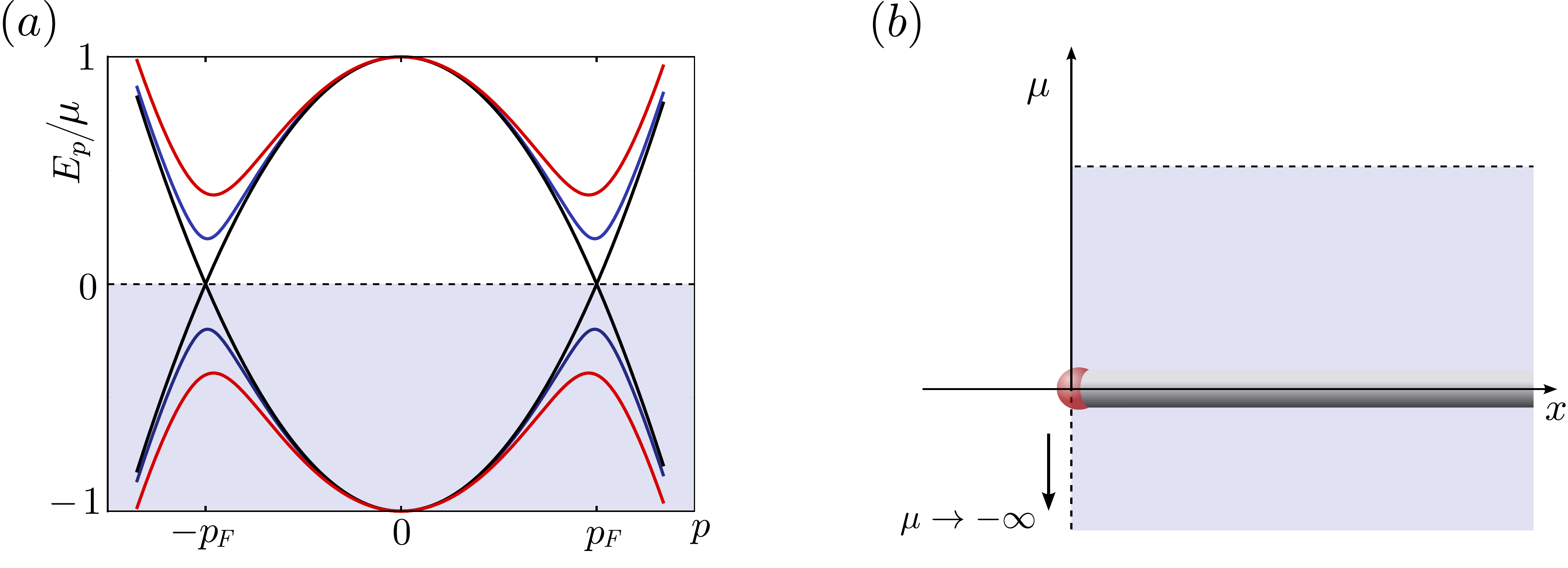}
  \caption{%
  (a) Plot of the energy-momentum relation of a \textit{p}-wave
  superconducting nanowire. All energies are measured with respect
  to the chemical potential $\mu >0 $ which is indicated by the dashed
  line. Above the chemical the spectrum corresponds to an excitation
  spectrum; below the chemical potential it should be understood as an
  absorption spectrum. Due to the presence of the superconductivity the
  two spectra are equivalent when mirrored at the dashed line. The black
  line shows two parabolas which is the spectrum without superconductivity,
  i.e., $\Delta=0$.  The blue and red lines correspond to a superconducting
  pair-potential $\Delta = 0.2 \mu$ and $\Delta= 0.4\mu$, respectively. The
  presence of the superconductor opens a gap of size $2\Delta$ at the
  chemical potential. Note that for any finite superconducting pairing
  $\Delta\neq0$ the system is completely gapped.  At $p=0$ and $E_{p=0}
  =\pm\mu$ the bottom of the band is visible. Decreasing $\mu$ brings the
  two vertices of the black parabolas closer together until they cross
  for $\mu=0$ which is the point of the topological phase transition. (b)
  Interface of a \textit{p}-wave superconducting nanowire (gray cylinder)
  with vacuum. The interface is located at $x=0$. To the right there is
  the nanowire with $\mu>0$ the vacuum is modeled by $\mu\to-\infty$, i.e.,
  a vanishing electron density. Due to the fact that the topological charge
  changes at the interface a single zero mode emerges which is a Majorana
  fermion depicted by the red sphere.
  }\label{fig:spectrum}
\end{figure}
Being interested in solutions of \eqref{eq:h_BdG_p_wave} with vanishing
eigenvalue, we note first that the spectrum of the system $E_p =
\sqrt{\xi(p)^2 + \Delta^2 (p/p_F)^2}$ is fully gapped in the translationally
invariant case with $\mu\neq 0$, see Fig.~\ref{fig:spectrum}(a). In order to
find a zero mode, we have to look at interfaces between different materials
such that the parameters $\mu, \Delta, m$ become spatially dependent. In the
simplest case, we look at an interface between vacuum for $z<0$ (modeled
by $\mu \to -\infty$) and a \emph{p}-wave superconductor for $z>0$, see
Fig.~\ref{fig:spectrum}(b). To simplify the discussion, we assume that only
the states close to the Fermi surface are important\footnote{Physically
this assumption means that the interface is not too abrupt in order not to
scatter states with vastly different momentum into each other.} and that the
wire is long enough such that the second interface does not influence
our discussion.  Thus, we can write $\psi(z) = e^{i k_F z} \psi_L(z) +
e^{-i k_F z} \psi_R(z)$ where we assume $\psi_L$ and $\psi_R$ to be slowly
varying on the scale $\lambda_F$.  This ansatz effectively linearize the
spectrum $\xi(p)$ around the Fermi points and leads to the Hamiltonian
\begin{equation}\label{eq:h_linearized}
  h_\text{BdG} = v_F p \,\eta^z +\Delta \,\tau^y\eta^y
\end{equation}
 with $v_F = p_F/m$ and where $\eta^z$ acts on the left/right-moving basis.
As the Hamiltonian commutes with $\tau^z \eta^z$, we can block diagonalize
it in each of the eigenspaces $\tau^z \eta^z = \pm 1$ with the result of
having two decoupled problems. A state $v_0$ at zero energy thus satisfies
the differential equation $h_\text{BdG} \,v_0 = 0$ with $p=-i\hbar\partial_z$.
The solutions are given by
\begin{align}\label{eq:solution}
  (v^\pm_0)_{\tau^z,\eta^z} = 
  e^{-z/\xi- i \pi \tau^z/4} \,
  \delta_{\tau^z,\pm \eta^z}
\end{align}
with the superconducting coherence length $\xi = \hbar v_F/\Delta$;
note that we have retained only those solutions which are exponentially
decaying away from the interface such that the wavefunction remain
normalizable. Additionally, we have chosen the overall phase of the
wavefunctions such that $\mathcal{C} v_0^\pm = v_0^\mp$.

The system is terminated with vacuum for $z<0$. In our description in
terms of linearized spectrum, the boundary condition with vacuum is
implemented by demanding a vanishing quasiparticle density at $z=0$. The
quasiparticle density operator is given by $\rho = \tfrac12 ( 1 + \eta^x)
\delta(z)$. Thus, we seek a solution of the form $v_0(z) = \alpha v_0^+(z)
+ \beta v_0^-(z)$ with the requirement $0 \stackrel!= \langle v_0|\rho|
v_0\rangle \propto |\alpha+ \beta|^2$. The condition is satisfied for
$\alpha=-\beta$ yielding a single bound state of the form
\begin{equation}\label{eq:def_v0}
  v_0(z) =  \alpha e^{-z/\xi} \begin{pmatrix}e^{-i\pi/4} \\
    - e^{i\pi/4} \end{pmatrix}_{\makebox[0pt]{$\scriptstyle\tau$}}
    \otimes \begin{pmatrix}1\\
    - 1 \end{pmatrix}_\eta
\end{equation}
at zero energy located at the interface between the \emph{p}-wave
superconductor and vacuum. In order that the Bogoliubov operator
$\gamma_1=\beta_0$ associated to $v_0$ is a Majorana operator with
$\beta_0^\dag=\beta_0$, we need to have $\mathcal{C} v_0 = v_0$ which demands
that $\alpha = \pm i \mathbb{R}$.\footnote{We do not allow for $\alpha\in
\mathbb{C}$ as the corresponding operator $\beta_0$ would not be Hermitian
in this case. What is left is the possibility to choose the sign of $\alpha$
which exactly corresponds to the $\mathbb{Z}_2$ symmetry described above.} The
proper normalization of the Majorana operator to $\gamma_1^2=1$ is realized
with $\alpha = i/\sqrt{2 \xi}$ and the Majorana operator assumes the form
\begin{align}\label{eq:maj_operator}
  \gamma_1&= \int_{z\geq 0}\!dz \, v_0(z)^\dag \Psi(z) 
  = (2\xi)^{-1/2}\! \int_{z\geq 0}\!dz\,
  e^{-z/\xi} \left\{e^{i\pi/4} [\psi_L(z) - \psi_R(z)]
  + \text{H.c.} \right\} \nonumber\\
  &= \sqrt{\frac2\xi} \int_{z\geq 0}\!dz\, \sin(k_F z) e^{-z/\xi} 
  \left[e^{-i\pi/4}\psi(z)  + e^{i\pi/4}\psi^\dag(z) \right].
\end{align}

The analysis can be generalized to include the second interface of the
\emph{p}-wave superconducting nanowire with vacuum which yields a second
Majorana operator $\gamma_2$.  In fact, it is a physical constraint that
on each connected piece of superconductor there is an even number of
Majorana fermions. Two Majorana fermions together can host a conventional
fermionic quasiparticle. Note that due to the superconducting condensate
the fermion number is not conserved.\footnote{Two electrons can be
taken out of superconducting condensate by breaking up a Cooper-pair.}
However, the fermion parity is still conserved. Thus in the situation as in
Fig.~\ref{fig:spectrum}(b), the parity $\mathcal{P} = - i\gamma_1 \gamma_2$
encodes the parity of the total number of electrons in the combined system
of the superconducting island and the nanowire.

\subsection{Topological charge}\label{sec:top_charge}

The appearance of the single zero mode which lead to the Majorana operator
in the last section was not an accident. In fact, its presence follows from
the general classification of topological insulators/superconductors that at
interfaces between superconductors with different topological charge there
will be a certain number of topological states which are protected and do
not depend on microscopic details. In the classification of topological
matter, one considers the question whether two Hamiltonians can be smoothly
deformed into each other without closing the gap. In the classification of
topological superconductors, one does not allow for arbitrary single-particle
Hamiltonians but restricts the class of Hamiltonians in such a way that one
only allows for Hermitian matrices which have a particle-hole symmetry and
thus represent a Bogoliubov-de Gennes Hamiltonian of a mean-field problem of
the form Eq.~\eqref{eq:nambu} \cite{schnyder:08,kitaev:09}. A topological
charge is an integer which is the same for all Hamiltonians which can be
deformed into each other and which differs for two Hamiltonian for which
this cannot be achieved. The classification tells us that in the case of
a one-dimensional superconductor with particle-hole symmetry $\mathcal{C}$
and $\mathcal{C}^2=1$ (as in our case), there are two distinct topological
classes. We have found that at an interface between vacuum (read conventional
insulator) with $\mu\to-\infty$ and a \emph{p}-wave superconductor there
is always a bound state \cite{teo:10,silaev:10,fukui:10}. This indicates
that the two parts are in different topological classes; we denote the
topological charge $\mathcal{Q}$ of the trivial insulator by $0$ and the
one of the \emph{p}-wave superconductor by $1$. As the topological charge
can only change when the gap closes and the gap of \eqref{eq:h_BdG_p_wave}
only closes for $\mu=0$, we find that
\begin{equation}\label{eq:hbdg}
  \mathcal{Q}(h_\text{BdG}) = \begin{cases} 1, & \mu > 0, \\
    0, & \mu<0.\end{cases}
\end{equation}
Due to this reasoning, the Majorana mode is always present in the
model as long as $\mu$ changes sign across the interface. For a more
complete discussion of this fact in the simpler model of polyacetylene
see Ref.~\cite{jackiw:81}.

\section{Quantum computation with Majorana fermions}

If we think about an implementation for a quantum computer, we are used
to the example of a spin-$\tfrac12$ particle which is a drosophila for a generic
two-level system \cite{feynman:3}. However, we can ask ourself the question
whether we can also use the many-body Fock space for quantum computation
purposes. We know that the occupation states $|n_1,n_2, \dots, n_N \rangle$
with $n_j\in\{0,1\}$ form a basis for the $N$-mode fermionic Fock space
generated by the creation operators $c^\dag_j$, $j\in\{1,\dots,N\}$,
starting from the vacuum state denoted by $|0\rangle$. The Fock space has
dimension $2^N$ (each mode can be either occupied or empty). Thus counting
the degrees of freedom, we are tempted to conclude that a fermionic system
with $N$-modes emulates $N$-qubits. In the next section, we will see that
this na\"ive counting argument is not completely correct as it violates
the so-called superselection rule. We will argue that quantum computation
with noninteracting fermions is not complete and will show what is needed
to make the setup complete. Then, we will show that Majorana fermions are
in fact non-Abelian particles such that some gates can be performed in a
parity-protected way.

\subsection{Fermionic quantum computation}\label{sec:fermion_qc}

Expressing a Hamiltonian $H$ or in fact any physical observable $A$ which
are bosonic operators in terms of fermionic creation and annihilation
operators, we are bound to only include terms where an even number of
fermion operators appear.\footnote{From the correspondence principle, we
know that for large quantum numbers the expectation values of operators
for physical observables should behave like (real) numbers. Due to the
anticommutation relation of fermionic operators, the correspondence
principle for a potential fermionic observable would instead lead to
anticommuting Gra{\ss}mann numbers.} The result is that the total fermion
parity $\mathcal{P} = \prod_{k} \mathcal{P}_k = (-1)^{\sum_k n_k}$ is
strictly conserved in a closed system; the reason for this is the fact that
\begin{equation}\label{eq:parity_on_hamilton}
  \mathcal{P} A \mathcal{P} =A
\end{equation}
which follows from $\mathcal{P} c_j \mathcal{P} = -c_j$ and the fact
that each term in $A$ involves an even number of fermionic operators.
Note that the superconducting Hamiltonian \eqref{eq:ham_eff} conserves the
total fermion parity even so the number of fermions is not conserved. Due
to this constraint, we have the following superselection rule: given two
states in a fermionic Fock space $|\psi_+\rangle$ and $|\psi_-\rangle$ with
different fermion parity, $\mathcal{P} |\psi_\pm\rangle = \pm |\psi_\pm
\rangle$ we have
\begin{equation}\label{eq:superselection}
  \langle \psi_- | A | \psi_+ \rangle =\langle \psi_- | \mathcal{P} A
  \mathcal{P}| \psi_+ \rangle = - \langle \psi_- | A | \psi_+ \rangle =  0
\end{equation}
for all observables $A$.  Thus, there is no point in making superpositions
between states of different parity as there will be no effect on any
observable. We can thus restrict ourselves to one superselection sector
and keep the fermion parity fixed with either $\mathcal{P} = +1$ or
$\mathcal{P}=-1$. The conclusion of this argument is that out of the $2^N$
states in a fermionic Fock space, only $2^{N-1}$ can be effectively used
for quantum computation purposes.

A further restriction to quantum computation using fermions arises
from the fact that noninteracting fermions subject to beam splitters,
phase-shifters (delay lines), measurements of the state of a single
electron (so-called fermionic linear optics) does in fact not lead to any
entanglement \cite{terhal:02}. In order to generate entanglement, we need
to add parity measurement of two electrons which effectively involves
interactions between different electrons \cite{beenakker:04b}.

\subsection{Parity-protected quantum computation}

\begin{figure}[tb]
  \centering
  \includegraphics[width=0.7\linewidth]{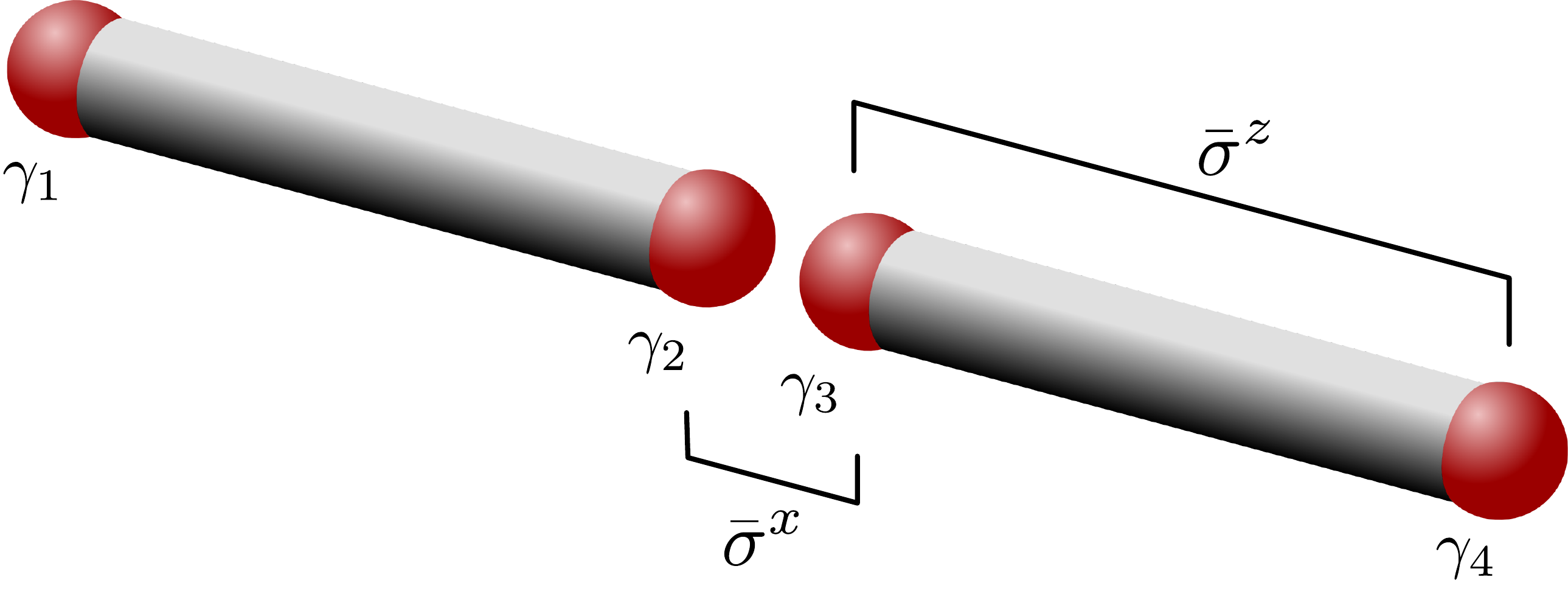}
  \caption{%
  Sketch of the parity Majorana qubit. Two Majorana fermions together form a
  single Dirac fermionic mode whose Hilbert space is two-dimensional as the
  mode can either be empty or filled, both states at the same energy. Four
  Majorana fermions thus form a four-dimensional Hilbert space of which due
  to the conservation of the total fermion parity only a two-dimensional
  subspace can be accessed. This degenerate two-dimensional subspace is the
  Majorana qubit. Gates on the qubit can be either performed by braiding or
  by coupling two Majorana fermions. As indicated in the figure, coupling
  $\gamma_3$ to $\gamma_4$ implements a $\bar \sigma^z$-operation whereas
  coupling $\gamma_2$ to $\gamma_3$ leads to a $\bar \sigma^x$-operation.
  Given the fact that the Majorana fermions are sufficiently far apart from
  each other and that the environment only acts locally on the system,
  these operations are not performed `accidentally' by the environment
  and the Majorana qubit is protected from both sign flip and bit flip
  errors. As these protection originates from the conservation of the
  total fermion parity, the qubit is called parity-protected.
  }\label{fig:parity_qubit}
\end{figure}
We have seen in the last section that due to the parity-conservation,
we need to have two fermionic modes to encode a single qubit. For
concreteness, we will work in the even parity superselection sector and
have the single logical qubit encodes as $|\bar 0 \rangle=|00\rangle$ and
$|\bar 1\rangle=|11\rangle$. Thinking about a possible implementation
in terms of Majorana fermions, we encode each fermionic mode in a
pair of Majorana fermions which are localized states sufficiently
far separated from each other. As we have seen before, two segments of
\emph{p}-wave superconducting nanowires exactly implement this situation,
see Fig.~\ref{fig:parity_qubit}. We denote the Majorana fermions in the
left segment as $\gamma_{1}$ and $\gamma_{2}$ and the one on the right
segment as $\gamma_{3}$ and $\gamma_{4}$ correspondingly. The Majorana
fermions are at zero energy thus the two states $|\bar 0 \rangle$ and
$|\bar 1 \rangle$ are degenerate in energy. The parity of the number of
electrons on the superconducting segments are given by $\mathcal{P}_{L}
= -i \gamma_{1}\gamma_{2}$ and $\mathcal{P}_R= -i \gamma_3 \gamma_4$. Due
to the parity constraint, we have $\mathcal{P}_L = \mathcal{P}_R$ and
the action of both operators on the logical qubit emulates the $\sigma^z$
Pauli-operator,
\begin{equation}\label{eq:sigma_z}
  \bar\sigma^z = - i \gamma_{1}\gamma_{2} =  - i \gamma_{3}\gamma_{4}
\end{equation}

In order to have a complete qubit, we are left with the task to find
a logical $\bar\sigma^x$, an operator which anticommutes with $\bar\sigma^z$.
It is easy to see that
\begin{equation}\label{eq:sigma_x} 
  \bar \sigma^x =-i \gamma_{2} \gamma_{3}
  =- i \gamma_{1} \gamma_{4}
\end{equation}
anticommutes with $\bar\sigma^z$ due to the fact that the single Majorana
fermions shared by both operators anticommute with each other. In
the situation where all the Majorana fermions are sufficiently far
separated from each other, either gate on the logical qubit is a nonlocal
operator. Due to this nonlocality, it is highly unlikely that uncontrolled,
random fluctuations in the environment will execute a gate thus acts as
an error on the logical qubit. This protection of the Majorana qubit is
called symmetry-protected topological order \cite{gu:09,pollmann:12} or
simply parity-protection \cite{hassler:11}. The decisive difference to
full topological order, as it is for example present in Kitaev's toric
code \cite{kitaev:06}, is the fact that logical Pauli operators are only
required to be nonlocal as long as the parity symmetry is conserved. Having
a reservoir tunneling single electrons on the superconducting island is a
local process which violates the parity-conservation and immediately brings
the Majorana qubit out of its computational subspace. 

The requirement for operating the Majorana qubit successfully in a
protected manner is that the environment does not provide single unpaired
electrons.  This sounds on the first sight very stringent. However, the
physical implementation of the system does only involve superconductors
where most of the electrons are paired up into Cooper pairs and where at
temperature $T$ only a exponentially small fraction proportional to the
Boltzmann factor $e^{-\Delta/T}$ remains unpaired.  The storage time of
quantum information in a Majorana qubit thus will increase exponentially
when lowering the electron temperature.

\subsection{Anyons}

\begin{figure}[tb]
  \centering
  \includegraphics[width=0.9\linewidth]{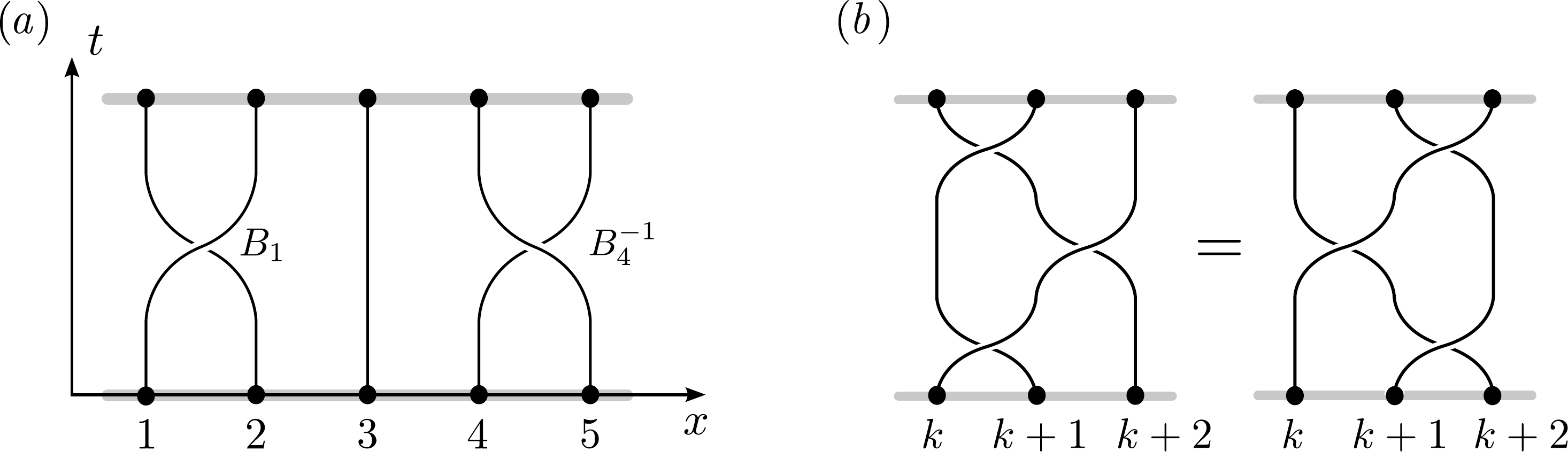}
  \caption{%
  (a) Elementary operation of the braid group. The geometric representation
  of the braid group is in space-time; the horizontal axis is the spacial
  axis whereas the vertical one is temporal. Every element (called braid) of
  the braid group $\mathcal{B}_N$ consists of $N$ strands. In our example,
  we have $N=5$ strands which are numbered from 1 to 5. The generators
  of the group $B_k$ denote the braiding two strands counterclockwise (in
  the figure, $B_1$ braids strand 1 and 2 counterclockwise). The inverse
  operation braids the strand clockwise (in the figure, $B_4^{-1}$ braids
  strand 4 and 5 clockwise). Two elements of the group are equivalent if
  the corresponding braids can be smoothly deformed into each other without
  moving the endpoints denoted by the black dots. The Yang-Baxter equation
  $B_k B_{k+1} B_k = B_{k+1} B_k B_{k+1}$ provides an important relation
  between the generators $B_k$ and $B_{k+1}$ and is shown pictorially in
  (b). That the two braids are topological equivalent can be seen as follows:
  in both braids strand $k+2$ can be considered to lie in the very back
  and to end up at the initial position of strand $k$. Similarly strand
  $k$ lies in front and ends up at the initial position of strand $k+2$.
  The middle strand starts at $k+1$ and ends at the same place. The braids
  are equivalent as they can be deformed into each other by sliding the
  middle strand $k+1$ in between the two other strands from the left to
  the right.
  }\label{fig:braiding}
\end{figure}
In $3+1$ dimensions, we are use to the dichotomy of bosons and fermions. In
fact the spin-statistic theorem can be proven in the context of relativistic
field theory which states that particles with integer spin are bosons
whereas particles with half-integer spin are fermions. The origin of this
distinction lies in the fact that the Hamiltonian of identical particles
commutes with an arbitrary element of the symmetric group $\mathcal{S}_N$
which exchanges the $N$ identical particles. Thus, it is possible to classify
the eigenstates of the Hamiltonian in terms of irreducible representations
of the permutation group. Any representation whose dimension is larger
than one leads to a degeneracy, which is called exchange degeneracy as it
originates simply from the fact that particles are indistinguishable. Now
it is an experimental fact that exchange degeneracies do not exist; the
absence of exchange degeneracy was first noted in the context of statistical
mechanics where it manifests itself in an entropy which is not extensive and
where it has been dubbed Gibbs paradox.\footnote{The statistics of identical
particles which transforms according to higher dimension representations of
the permutation groups is called parastatistics.  However, even if particles
with parastatistics where to exist they would offer nothing new as a set of
Klein transformations could be used to map particles with parastatistics
as bosons or fermions with a set of internal quantum numbers (like spin,
\dots). Later, we will see that such a mapping is not possible in $2+1$
dimension and that higher dimensional representations of the braid group
are truly different from the one-dimensional representations.}

In $2+1$ dimension, the relevant group is the braid group $\mathcal{B}_N$ of $N$
strands as trajectory in space-time for exchanging two particles clock or
counterclockwise are topological distinct.\footnote{In $3+1$ dimension
clock and counterclockwise depends on the observer (coordinate system)
and thus the two exchanges are topologically equivalent.} We denote
with $B_{j}$ the counter-clockwise exchange of strand $j$ and $j+1$
($1\leq j\leq N-1$). Note that different from the symmetric group $B_j
\neq B_j^{-1}$. The braid group fulfills the following relations
\begin{equation}\label{eq:braid_group}
  B_k B_l = B_l B_k, \quad |k-l|\geq 2
  \qquad\text{and}\qquad
  B_k B_{k+1} B_k = B_{k+1} B_k B_{k+1},
\end{equation}
the latter is called Yang-Baxter equation.  Different from the symmetric
group $\mathcal{S}_N$ the group order is infinity which makes the
classification of all irreducible representation difficult.

The one-dimensional (unitary) representations of the braid group are
simple to construct; representing the action of $B_j$ onto a wavefunction
by $e^{i\theta_j}$ with $\theta_j \in [0,2\pi)$, we immediately get from
the Yang-Baxter equation that all the angles are equal, i.e., $\theta_j
\equiv \theta$. Note that for $\theta=0$, we get the customary result for
bosons that interchanging two particles does nothing to the wavefunction
whereas for $\theta=\pi$ interchanging introduces a minus sign which
is the result for fermions. In $2+1$ dimension, all angles in between
0 and $\pi$ are allowed and particles with $\theta \neq 0$ or $\pi$ are
called (Abelian) anyons. As an example, we note that quasiparticles in
the fractional quantum Hall effect at filling fraction $\nu = \tfrac1n$
with $n$ an odd integer are anyons with $\theta = \nu \pi$.

Particles whose wavefunctions transform according to higher dimensional
irreducible representations of the braid group are called non-Abelian
anyons. A necessary ingredient is a ground state degeneracy (which grows
exponentially with the number of particles). The effect of $B_j$, the
counterclockwise exchange two particles $j$ and $j+1$, is then represented
by a unitary matrix $U_j$ on the ground state manifold. As different
unitary matrices do not commute, the representation is non-Abelian and
thus the particles are called non-Abelian anyons.  The usefulness of
non-Abelian anyons for topological quantum computation relies on the fact
that the degeneracy of the ground state manifold is protected and the gates
implemented by the exchange of particles are exact (up to an unimportant
global phase) \cite{nayak:08}. If for a specific species of non-Abelian
anyons for any given gate a braid can be found which approximates the gate
with arbitrary accuracy, the non-Abelian anyons are called universal for
quantum computation.

\subsection{Majorana fermions as
non-Abelian particles}

\begin{figure}[tb]
  \centering
  \includegraphics[width=0.9\linewidth]{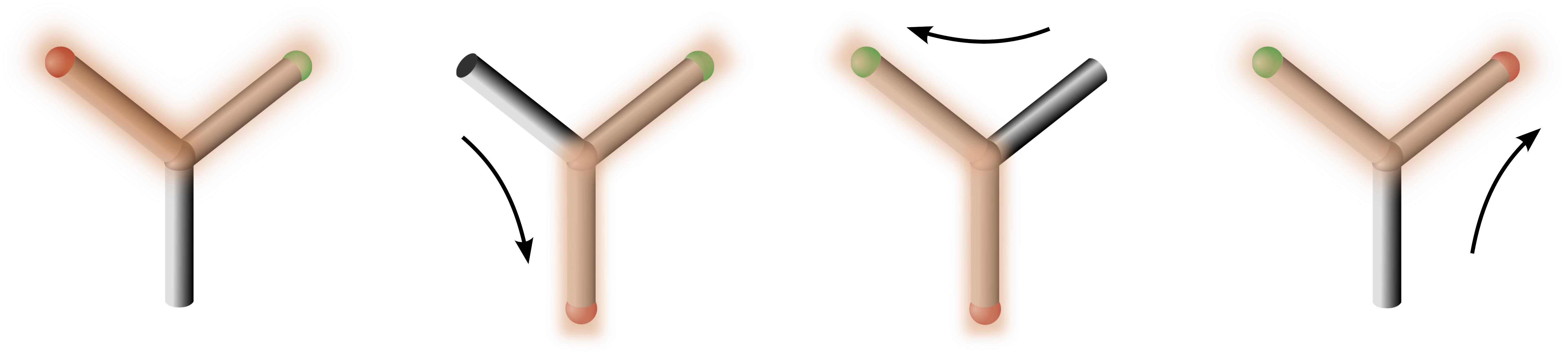}
  \caption{%
  Sketch of a Y-junction where three superconducting nanowires meet. Having
  a Y-junction is an essential ingredient to be able to braid the Majorana
  fermions as exchanging two particles is not possible in a strictly
  one-dimensional setting. The two Majorana fermions symbolized by red and
  green spheres are situated at the ends of the segment of the wire in the
  topological phase with $\mathcal{Q}=1$ indicated by the brown shading. The
  gray part of the nanowire is depleted such that it is topological trivial
  with $\mathcal{Q}=0$. Exchanging the red and green Majorana fermion is
  done in four steps indicated in the figure. In each step, one of the
  Majorana fermion is moved to another end of the Y-junction by swapping
  the corresponding segments of the wire between topological and trivial.
  }\label{fig:y_junction}
\end{figure}

We restrict ourself to the discussion of two Majorana fermions. Braiding is
local, so it should only affect those particles which are being braided. We
will see that the braid statistics can be simply deduced from the fact that
the parity remains conserved \cite{halperin:12}.\footnote{In two dimensions,
Majorana fermions exist as bound state in the vortex of chiral \emph{p}-wave
superconductors. To determine the braid statistics in this case is a bit
more involved \cite{read:00,ivanov:01}.} First, we have to see that moving
a Majorana fermions: braiding of the Majorana fermions the system has to
stay in the ground state manifold. Thus everything has to be adiabatic. We
change parameters in the Hamiltonian slowly in such a way that everything
is slow with respect to the gap. The system then evolves according to the
unitary evolution $U(t) = \mathcal{T} \exp[-i\int_0^t dt'\, H(t')]$ where
$\mathcal{T}$ is the time ordering operator. The operators transform like
\begin{equation}\label{eq:time_evol}
 \gamma_k(t) = U^\dag(t) \gamma_k U(t).
\end{equation}

In a first step, we want to show that we can in fact move a single
Majorana fermion without changing its state.  Having two adjacent Majorana
fermions $\gamma_k$ and $\gamma_{k+1}$ moving the former a bit in the
time $T$ results in new Majorana fermions $\gamma_k'=\gamma_k(T)$ and
$\gamma_{k+1}'=\gamma_{k+1}(T)$. As fermion parity is conserved, we have
\begin{equation}\label{eq:parity_conserved}
  \mathcal{P} = - i \gamma_k \gamma_{k+1} = -i \gamma_k'\gamma_{k+1}'
\end{equation}
Since nothing has happened to $\gamma_{k+1}$ (the evolution did not affect
the right Majorana fermion), we have $\gamma_{k+1}'=\gamma_{k+1}$. Plugging
this into Eq.~\eqref{eq:parity_conserved}, we have $\gamma_k'=\gamma_k$,
i.e., we can move the Majorana fermion without perturbing it.

In order to obtain the braid group, we need to find the effect of $B_k$,
i.e., exchanging the two Majorana fermions in the clockwise direction. This
cannot be done in strictly one dimension but involves $Y$-junctions, see
Fig.~\ref{fig:y_junction} and Refs.~\cite{alicea:11,clarke:11,heck:12}. As
above, we will again denote the operators after time $T$, that is after
the braiding operators with a prime. As the positions of the Majorana
fermions is switched after the time $T$, we have $\gamma_k' = \alpha_{k+1}
\gamma_{k+1}$ and $\gamma_{k+1}' = \alpha_k \gamma_k$. Due to the fact
that the operators have to remain Majorana fermions, we need to require
that $\alpha_k,\alpha_{k+1} \in \{\pm 1\}$. The conservation of the parity
leads to the relation
\begin{equation}\label{eq:parity_conserved_2}
  \mathcal{P} = -i \gamma_k \gamma_{k+1} = -i \gamma_k'\gamma_{k+1}'
  = -i \alpha_k \alpha_{k+1} \gamma_{k+1} \gamma_k
\end{equation}
thus, we need that $\alpha_k \alpha_{k+1} = -1$: one of the Majorana fermions
picks up a minus sign and one does not. Which of the Majorana fermions picks
up a minus sign is a gauge choice and there are no physical effect; so we
choose $\alpha_k=-1$ and $\alpha_{k+1}=1$. Note that braiding the Majorana
fermions in the counter-clockwise direction is the inverse operation and
thus reverses the sign of both $\alpha_k$ and $\alpha_{k+1}$. It is easy
to see that the operator
\begin{equation}\label{eq:braiding}
  U(T) \equiv U_k = \exp \left( \frac\pi4 \gamma_k \gamma_{k+1}
  \right) = \frac{1}{\sqrt{2}} (1 + \gamma_k \gamma_{k+1}),
\end{equation}
implements this unitary operation, where $`\equiv'$ denotes the fact that
the time-evolution $U(T)$ and $U_k$ are equivalent up to an unimportant
overall phase. Furthermore, one can check explicitly that the $U_k$s
obey the relations in Eqs.~\eqref{eq:braid_group} and thus they are a
representation of the braid group.

We would like to explicitly find the gates generated by braiding the
Majorana zero mode for the example of a single Majorana qubit, cf.\
Fig.~\ref{fig:parity_qubit}. Straightforward calculation shows
\begin{align}\label{eq:braid_single_qubit}
  U_1 &= U_3= e^{\pi\gamma_{1}\gamma_{2}/4}  = e^{i \pi \bar\sigma^z/4},&
  U_2 &= e^{\pi\gamma_{2}\gamma_{3}/4}  = e^{i \pi \bar\sigma^x/4},
\end{align}
which are elements of the Clifford group. In fact, it can be shown that in
general only Clifford gates can be implemented via braiding. So neither
the single qubit rotations are universal nor there is an entangling
gate.\footnote{Majorana fermions are in essence noninteracting particles
described by the quadratic Hamiltonian Eq.~\eqref{eq:nambu} and thus
the restrictions of Sec.~\ref{sec:fermion_qc} apply.} However in order
to make the setup universal, we need to subjoin only a two qubit
parity measurement (to generate entanglement) and a $\tfrac\pi8$-phase
gate \cite{bravyi:05,nayak:08}.\footnote{In fact, it is enough if the
$\tfrac\pi8$-phase gate is implemented with a fidelity $\gtrsim 90\%$ as a
distillation protocol using the exact Clifford gates called the Magic state
distillation can be employed to purify the state \cite{bravyi:05}.} We will
show a way to implement such a unprotected gate in Sec.~\ref{sec:hybrid}.

\section{Implementations}

\subsection{Semiconducting nanowires}

Last year at TU Delft evidence of Majorana fermions has been found in
semiconducting InSb nanowires coupled to a NbTiN superconductor in the
presence of a magnetic field of around 100$\,$mT \cite{mourik:12}. The
experiment realized an idea of Refs.~\cite{lutchyn:10,oreg:10} which extended
earlier ideas for quantum wells to nanowires \cite{sau:10,alicea:10}. In this
chapter, we provide a brief introduction into the physics of semiconducting
nanowires and motivate the connection to Majorana fermions in spinless
\emph{p}-wave nanowires. The simplest model for conduction electrons a
semiconducting nanowire in a magnetic field $B$ pointed along the nanowire
direction is given by
\begin{equation}\label{eq:ham_snw}
  H_\text{NW} = \int\!dz\, \left[\frac{\hbar^2}{2m} |\Psi'(z)|^2 - \mu
  |\Psi(z)|^2  + \frac{\alpha_\text{so}}\hbar \Psi^\dag(z) \sigma^y p \Psi(z) 
  - \frac{g \mu_B B}{2} \Psi^\dag(z) \sigma^z \Psi(z)\right]
\end{equation}
with the field operator $\Psi^T = \begin{pmatrix}\psi_\uparrow
 &\psi_\downarrow \end{pmatrix}$. Here, we have assumed that the spin-orbit
 characterized by $\alpha_\text{so}$ is of Rashba-type
due to electric fields perpendicular to the substrate as Dresselhaus terms
are absent for experimentally-relevant zincblende nanowires grown along
the 111 crystal direction. For InSb nanowires the parameters are given by
$m\simeq0.015 m_e$, $g\simeq50$, and $\alpha \simeq 0.2\,{\rm eV}{\rm \AA}$.
In writing down the Hamiltonian \eqref{eq:ham_snw}, we have silently
assumed that only a single mode is relevant in the nanowire whereas in
the experiment there are most likely of the order of 5 modes contributing.
This difference, even though important for a detailed description of the
experimental findings, is not relevant for the generic discussion we intend
to provide. In the same spirit, we have neglected any orbital effects of
the magnetic field on the nanowire.

The proximity to an \emph{s}-wave superconductor introduces pair-correlations
in the nanowire. This can be expressed with an additional term in the
Hamiltonian of the form
\begin{equation}\label{eq:ham_sc}
  H_\text{SC} =  \Delta
  \int\!dz\, \left[
  \psi^\dag_\downarrow(z) \psi^\dag_\uparrow(z)
  + \psi^\pdag_\uparrow(z) \psi^\pdag_\downarrow(z) \right]
\end{equation}
where $\Delta/\hbar>0$ is the rate at which Cooper pairs are injected into
the nanowire. For reasonable strong superconductors and clean interfaces
between the superconductor and the nanowire we may expect $\Delta\simeq
1\,{\rm K}$.

Following Ref.~\cite{alicea:10}, we want to show that for strong magnetic
fields such that the Zeeman energy $E_Z = \tfrac12 g \mu_B B$ is larger than
$\Delta$ and $m\alpha_\text{so}^2/\hbar^2$ the system is equivalent to the
spinless nanowire discussed in Sec.~\ref{sec:spinless}. For such large
magnetic fields basically only electrons with spin-up are present. We
thus write $\psi(z) = \psi_\uparrow(z)$. As the spin-orbit term in
\eqref{eq:ham_snw} does not commute with the Zeeman term, it admixes the
spin-down component. In lowest order perturbation theory, we thus have
\begin{equation}
  \psi_\downarrow(z) = \frac{\alpha_\text{so}\sigma^y p}{2\hbar E_Z} \psi(z)
  = \frac{\alpha_\text{so}}{2\hbar E_Z} \psi'(z)
\end{equation}
Plugging this expression into the full Hamiltonian $H_\text{NW} +
H_\text{SC}$, we obtain to lowest order the effective Hamiltonian
\begin{equation}\label{eq:h_eff}
  H_\text{eff} = \int\!dz\,\left[ \frac{\hbar^2}{2m} |\psi'(z)|^2
   - \mu_\text{eff} |\psi(z)|^2
   - \Delta_\text{eff} [\psi^\dag(z) \psi^\dag{}'(z) -\psi(z)  \psi'(z)]\right]
\end{equation}
with $\mu_\text{eff} = \mu + E_Z$ and $\Delta_\text{eff} = \Delta
\alpha_\text{so}/2
\hbar E_Z$. The expression \eqref{eq:h_eff} coincides with the BCS mean-field
Hamiltonian $H_\text{MF}$ for spinless electrons Eq.~\eqref{eq:ham_eff}
with an effective \emph{p}-wave pairing $\Delta(z) = \Delta_\text{eff}
\delta'(z)$. We have seen in Sec.~\ref{sec:top_charge} that this system hosts
Majorana fermions at its ends provided that $\mu_\text{eff} >0$.

We know however that the topological phase is stable and can only be
removed by closing the bulk gap of the nanowire. The condition for the
closing of the gap of $H_\text{NW} + H_\text{SC}$ is a zero eigenvalue
of the Bogoliubov-de Gennes Hamiltonian
\begin{equation}\label{eq:h_bdg_nw}
  h_\text{BdG}(p) =  \xi(p) \tau^z + 
  \frac{\alpha_\text{so} p}{\hbar} \sigma^y  \tau^z
  -E_Z \sigma^z + \Delta \tau^x
\end{equation}
subject to periodic boundary conditions. Due to the presence of the spin-orbit
term proportional to $\alpha_\text{so}$ the gap can only be closed for $p=0$. At $p=0$,
the energy eigenvalues of $h_\text{BdG}$ are given by
\begin{equation}\label{eq:eigenvalue}
  E_{p=0} = \pm \left|\sqrt{\mu^2 + \Delta^2} - E_Z \right|.
\end{equation}
We have seen before that for large magnetic field with $E_Z \gg \Delta,|\mu|$
the system essentially implement a spinless \emph{p}-wave nanowire in the
topological phase with Majorana fermions at the end. This phase extends in the
full model up to the point where $E_{p=0}=0$, i.e., the gap closes.  As a
consequence, we obtain an expression for the topological charge
\begin{equation}\label{eq:top_charge}
  \mathcal{Q}(h_\text{BdG})
  =\begin{cases}
     1,& |\mu| < \mu_c ,\\
     0, &|\mu| > \mu_c,   \end{cases} 
     \qquad\text{with}
 \qquad \mu_c  =
   \begin{cases}  \sqrt{E_Z^2 -
\Delta^2},& E_Z > \Delta,\\
   0, & E_Z < \Delta.
\end{cases}
\end{equation}
In order for the experiments to be in the regime where there are
Majorana fermions (indicated by $\mathcal{Q} =1$), we thus need $E_Z >
\Delta$. This is only possible due to the large $g$-factor of the InSb
nanowire with respect to the superconductor with $g_\text{SC}\simeq 2$
as the superconducting state is destroyed latest at a magnetic field such
that $\tfrac12g_\text{SC} \mu_B B = (g_\text{SC}/g) E_Z \lesssim \Delta$
which is the so-called Pauli limit.

\subsection{Hybrid structures}\label{sec:hybrid}

As discussed in details above, the parity protection of the Majorana qubit
is potentially useful as it allows quantum computation where the qubit as
well as the gates are topological protected. However, we have also seen that
the braiding operation do not allow for universal quantum computation. In
particular, a $\tfrac\pi8$-phase gate and a parity measurement of two
qubits are missing. In this section, we want to show how hybrid structures
involving superconducting qubits besides the Majorana qubits offer
a way to make the Majorana qubits universal for quantum computation.

It is important to note that in order to be able to measure the state of
the Majorana qubit its parity-protection has to be removed.  This can in
principle be achieve by fusing two Majorana fermions, i.e., make their
wavefunctions overlapping, and measuring the resulting ground state
energy of the system. However, as we need additionally a measurement of
the parity of two qubits (without determining the state of each of the
qubits) interactions, terms which couple four Majorana fermions,
are indispensable \cite{nayak:08}.

\begin{figure}[tb]
  \centering
  \includegraphics[width=0.8\linewidth]{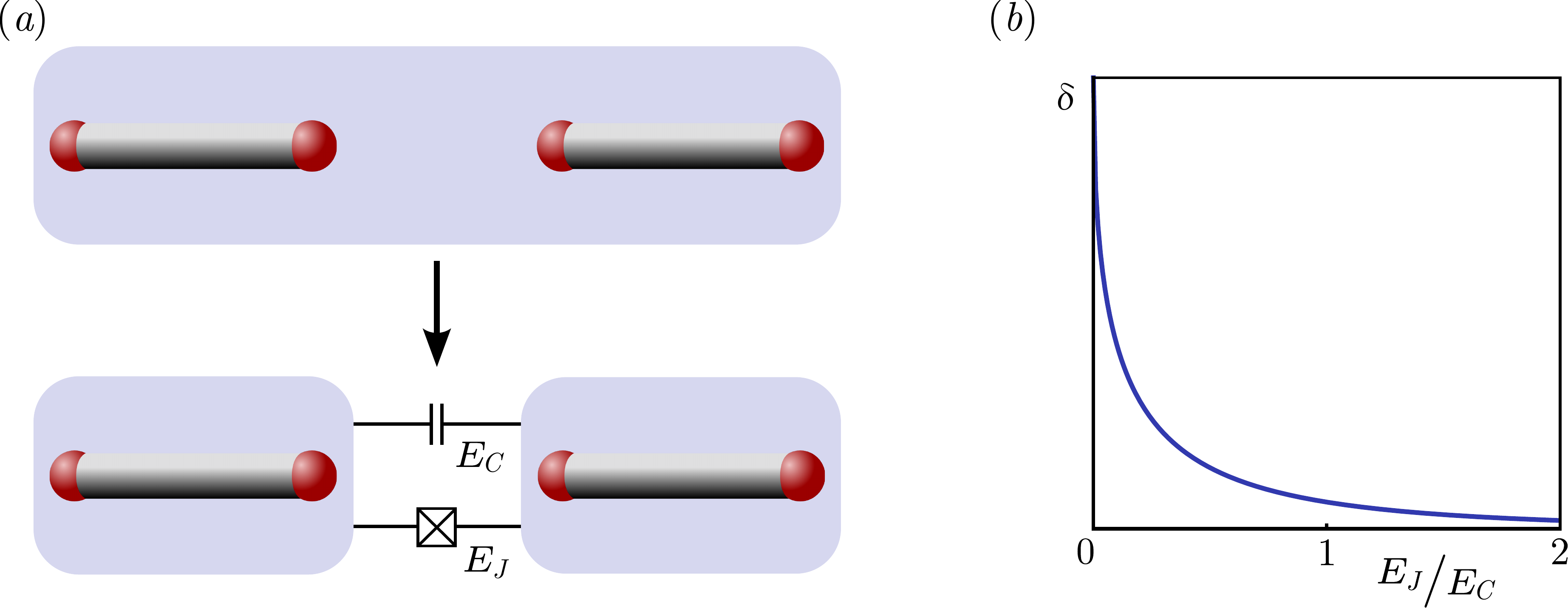}
  \caption{%
  (a) In order to measure and/or manipulate the Majorana qubit the
  topological protection has to be removed. This can be achieved by `cutting'
  the superconducting island into two parts, a process which is implemented
  in reality by having from the beginning two superconducting islands
  which are coupled by a capacitance with energy $E_C$ and a Josephson
  junction with energy $E_J$ (bottom). In the regime where $E_J \gg E_C$,
  superconducting correlations between the two islands are strong and tnus
  the two islands essentially behave like a single superconducting islands
  with a Majorana qubit consisting of four Majorana fermions. (b) Lowering
  the ratio $E_J/E_C$ the charge sensitivity $\delta$ which describes the
  energy difference between the two logical states of the Majorana qubit
  increases exponentially and readout and manipulation becomes possible.
  }\label{fig:measurement}
\end{figure}

The most natural choice for an interaction is the Coulomb interaction
due to the charge of the electrons which form the basis for the Majorana
fermions \cite{hassler:10c,hassler:11}. Starting with a Majorana qubit
involving four Majorana fermions on a superconducting island, cf.\
Fig.~\ref{fig:parity_qubit} the protection of the qubit can be lifted by
`cutting' the superconducting islands into two pieces each of them having
two Majorana fermions present. The state of the Majorana qubit is then
encoded in the fact whether there is an even or odd number of electrons
on each of the parts, see Fig.~\ref{fig:measurement}. The `cutting' is of
course not meant literally. In fact having from the beginning two separate
superconducting islands coupled with a Josephson coupling with strength $E_J$
(originating from Cooper pairs tunneling between the two superconductors)
and a capacitive coupling with strength $E_C$ (originating from the charging
energy between the two superconducting islands). In the regime $E_J \gg
E_C$, the two superconducting islands behave essentially as if they where
a single piece of superconductor and the qubit is protected. Lowering
the ratio $E_J/E_C$ increases the charge sensitivity (the difference in
energy between the islands having even and odd number of electrons) and
thereby lifts the parity-protection. In fact, the difference in energy
of the two states of the Majorana qubit is exponentially depending on the
ratio $E_J/E_C$ as $\delta \propto \exp\left(-\sqrt{8 E_J/E_C}\right)$,
see Ref.~\cite{koch:07}. The protected qubit discussed so far is given
for large $E_J/E_C$ where $\delta$  essentially vanishes.

\begin{figure}[tb]
\centering
\includegraphics[width=0.7\linewidth]{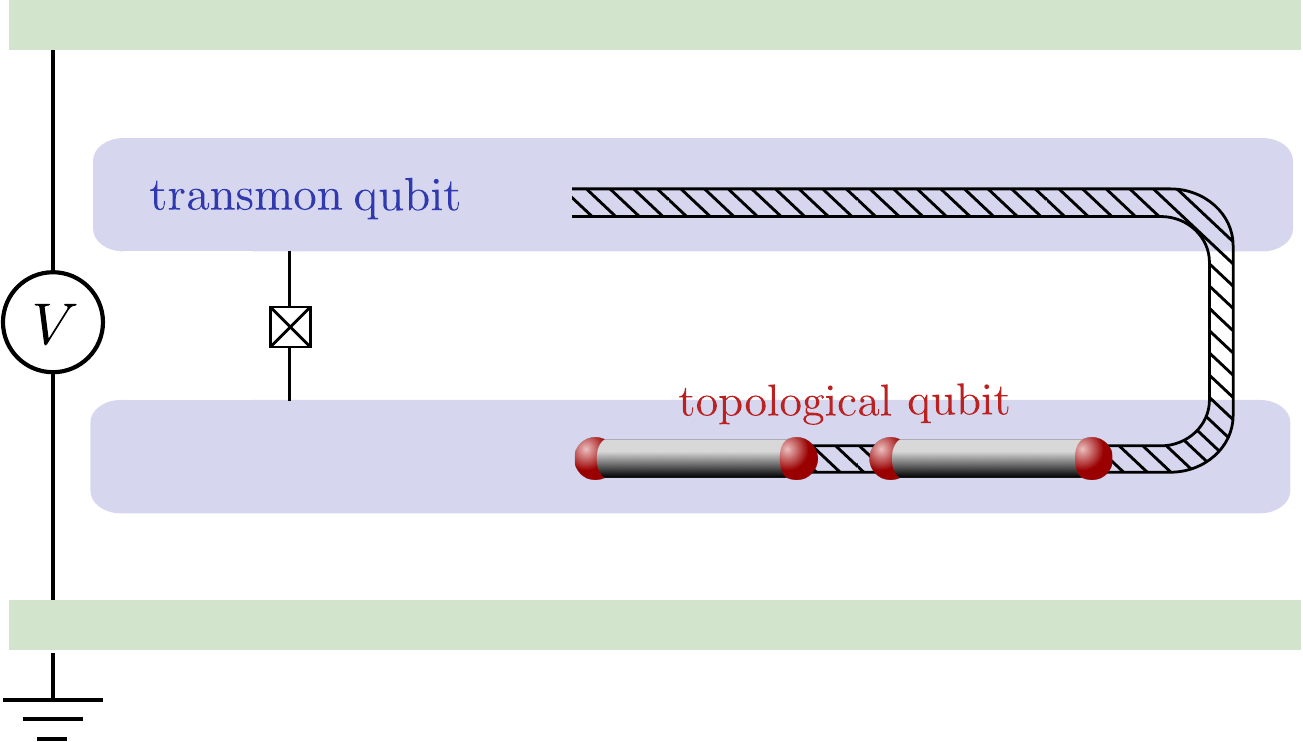}
\caption{%
Read out of a parity qubit in a Cooper pair box. Two superconducting islands
(blue), connected by a split Josephson junction (crosses) form the Cooper
pair box. The topological Majorana qubit is formed by four Majorana fermions
(red spheres), at the end points of two undepleted segments of a semiconductor
nanowire (striped ribbon indicates the depleted region). A magnetic flux
$\Phi$ enclosed by the Josephson junction controls the charge sensitivity
of the Cooper pair box. To read out the topological qubit, two of the four
Majorana fermions that encode the logical qubit are moved from one island to
the other. Depending on the quasiparticle parity, the resonance frequency in
a superconducting transmission line enclosing the Cooper pair box (green)
is shifted upwards or downwards by the amount which is exponentially small
in $E_J/E_C$.
}\label{fig:box}
\end{figure}

In the concrete implementation proposed in \cite{hassler:11}, the read-out
of the Majorana qubit is achieved by distribution the four Majorana fermions
which form the logical qubits on the two plates of the capacitor in a Cooper
pair box, see Fig.~\ref{fig:box}.  By varying the flux $\Phi$ through a split
Josephson junction, the Josephson energy $E_{J}\propto\cos(e\Phi/\hbar)$
becomes tunable.  As the charge sensitivity depends exponentially on the
ratio $E_J/E_C$, a variation of the charge sensitivity (the degree of
protection of the qubit) over two orders of magnitude has been achieved in
the transmon qubit design of the Yale group \cite{schreier:08}.\footnote{The
transmon is simply a Cooper pair box with $E_J \gtrsim E_C$ which is placed
in a transmission line resonator for read-out, hence the name.} Due to
this exponential dependence it becomes possible to turn the protection
of the qubit on and off at will.  The fact that there is a tunable energy
difference $\delta$ of the two states of the Majorana qubit directly leads
to the possibility of implementing an arbitrary phase gate. Starting and
ending with a qubit in the protected state with $\delta =0$, we can make
$\delta$ finite for some time $\tau$ such that $\delta \, \tau/\hbar =
2\phi$ which implement the phase gate $\exp(i\phi\bar\sigma_z)$. Especially
for $\phi=\tfrac\pi8$, we obtain the missing $\tfrac\pi8$ phase gate.

Similarly, for the readout of the Majorana qubit we use the fact that due to
the same interaction the energy difference $\Delta \varepsilon_{\bar\sigma_z}
= \Delta\varepsilon + \delta \,\bar\sigma_z$ between the ground state and
the first excited state of the transmon qubit depends on the state of the
Majorana qubit $\bar\sigma_z=\pm1$. The state of the transmon qubit can
be read out by sending a microwave probe beam through the transmission
line resonator \cite{wallraff:04,schuster:05,schreier:08}.  The resonance
frequency $\omega_\text{res}$ of the cavity is given by
\begin{equation}\label{eq:frequency}
  \omega_\text{
  res}=\omega_{0}-\frac{g^{2}}{\omega_{0}-\Delta\varepsilon_{\bar\sigma_z}/\hbar}
\end{equation}
provided the fact that the qubit is initially in its ground state and the
cavity frequency $\omega_0$ is far detuned from $\Delta \varepsilon/\hbar$;
here, $\omega_0$ is the bare resonance frequency of the cavity and $g$
is the Jaynes-Cummings coupling between the cavity mode and the transmon
qubit.  The small shift of the cavity frequency due to the dependence of
$\omega_\text{res}$ on the state $\bar\sigma_z$ of the Majorana qubit can
be measured sensitively as a phase shift of the transmitted microwaves.

In the case where there are more than two Majorana fermions per
superconducting island, the transmon qubit couples directly to parity
of the number of electrons on each of the island. Thus, a joint parity
measurement on two Majorana qubits can likewise be performed by moving
four out of the eight Majorana fermions to the other island.

Summarizing, the hybrid design of a coupled transmon and Majorana qubit
retains the full topological protection with exponential accuracy in the
off-state ($E_{J}/E_{C}\gg 1$). This hybrid device can be used to implement
a phase gate on the Majorana qubit as well as to jointly read out sets of
Majorana qubits. Together with braiding and single qubit readout, these
are the operations required for a universal quantum computer.

\section{Conclusion}

We have shown that Majorana fermions can be perceived as `half' and ordinary
Dirac fermion. Interestingly, these particles emerge as end states at
zero energy in superconducting \emph{p}-wave nanowires independent of
any microscopic details, a fact which can be traced back to the change
of the topological charge from the nanowire to the surrounding vacuum.
Four Majorana fermions encode a single qubit. Keeping the fermion parity in
the system conserved, gates on the qubit can only be performed by operators
which involve two Majorana fermions. As the Majorana fermions are spatially
separated, random (local) noise due to environmental fluctuations will not
couple to the Majorana qubit and thus will not lead to decoherence of the
Majorana qubit. Most importantly, we have seen that Majorana fermions are
non-Abelian particles which implies that one can perform protected gates
just by braiding the particles around each other. Alas, the operations
generated by braiding the particles are not enough to make the system a
universal quantum computer. For that reason, we have shown how the coupling
of a Majorana qubit to a superconducting transmon qubit can be used both for
performing measurments of the Majorana qubit and implementing the missing
gates.  Finally, we have discussed the essential ingredients (semiconducting
nanowire with strong spin-orbit interaction and large $g$-factor coupled
to an \emph{s}-wave superconductor) for implementing these ideas in a
laboratory without the need for the elusive \emph{p}-wave superconductor.


\end{document}